\documentclass{article}
\usepackage{spconf,amsmath,graphicx}
\usepackage{multirow, makecell}
\title{Audio Deepfake Detection with Self-Supervised WavLM and Multi-Fusion Attentive Classifier}
\name{Yinlin Guo, Haofan Huang, Xi Chen, He Zhao, Yuehai Wang}
\address{Zhejiang University\\
Department of Information and Electronic Engineering\\
HangZhou, China}
\begin{document}
%\ninept
%
\maketitle
\begin{abstract}
    With the rapid development of speech synthesis and voice conversion technologies,
    Audio Deepfake has become a serious threat to the Automatic Speaker Verification (ASV) system.
    Numerous countermeasures are proposed to detect this type of attack.
    In this paper, we report our efforts to combine the self-supervised WavLM model and Multi-Fusion Attentive classifier for audio deepfake detection.
    Our method exploits the WavLM model to extract features that are more conducive to spoofing detection for the first time.
    Then, we propose a novel Multi-Fusion Attentive (MFA) classifier based on the Attentive Statistics Pooling (ASP) layer.
    The MFA captures the complementary information of audio features at both time and layer levels.
    % The MFA focuses on the multi-layer representations extracted by the WavLM at both time and layer levels.
    % Experiments demonstrate that our proposed method reduces the equal error rate (EER) to 2.56\% on the ASVspoof 2021 DF evaluation set, outperforming current state-of-the-art anti-spoofing systems.
    % We further show the robustness of our model on the ASVspoof 2021 and 2019 LA evaluation set. 
    Experiments demonstrate that our methods achieve state-of-the-art results on the ASVspoof 2021 DF set and provide competitive results on the ASVspoof 2019 and 2021 LA set.
    
    % To our best knowledge, it is the highest reported performance on this dataset so far.
\end{abstract}

\section{Introduction}
\label{sec:intro}
% 先讲虚假语音检测

Automatic Speaker Verification (ASV) system is an important biometric solution widely used in identity authentication applications
such as access control systems, telephone banking, and forensic scenarios \cite{Anjum_2017}.
It operates by verifying the claimed speaker's identity through specific features derived from speech signals.
However, with the development of speech synthesis and voice conversion technologies,
a large number of spoofed audio samples have emerged that are difficult to distinguish from real samples \cite{Khan2022VoiceSC}.
Such samples can easily deceive both humans and ASV systems, posing significant challenges for speech anti-spoofing research.

% 再讲语音自监督模型
With the improving interest in developing robust spoofing countermeasures (CMs), self-supervised features have gained increasing attention in recent research.
Speech self-supervised models are capable of leveraging large amounts of unlabeled data to extract representations of speech signals.
Many researchers have already utilized these models as front-end feature extractors for audio deepfake detection.
Xie {et al.} \cite{xie21_interspeech} propose the utilization of Wav2vec2 model \cite{wav2vec2} to train an embedding siamese neural network to protect anti-spoofing models from attacks.
A similar method \cite{donas_icassp} uses the pre-trained Wav2vec2 \cite{wav2vec2} and a downstream classifier to detect spoofed audio. %  on the 2022 ADD challenge \cite{add_2022}
Wang {et al.} \cite{wang22_odyssey} explore different pre-trained self-supervised speech models, such as Wav2vec2 and XLS-R \cite{xlsr}, as the front-end of spoofing CMs.
They find that a self-supervised front-end pre-trained using diverse speech data performed quite well.
Recently, Tak {et al.} \cite{tak22_odyssey} also achieve significantly improved performance in the filed of spoofing detection by applying the XLS-R model.

% \cite{xlsr} with a self-attentive aggregation layer and data augmentation, their method 
% achieving significantly improved performance in spoofing detection.

However, features from current speech self-supervised models are inadequate for multi-speaker tasks \cite{chen2022wavlm}, including audio deepfake detection.
The models are usually trained with masking prediction pretext task.
During the pre-training stage, a certain percentage of time steps are masked in the latent feature encoder space.
The model is learned to identify the quantized latent audio representation for each masked frame.
Despite their excellent performance in tasks such as phoneme classification and automatic speech recognition, their effectiveness is constrained when it comes to certain speaker-related tasks.
Features retrieved by these models contain the content information of the audio samples,
but neglect speaker-related attributes which are more compatible with audio deepfake detection due to the presence of speaker-related artefacts in spoofing speech \cite{Anjum_2017}.

% 然后写，本文 balabala
In this paper, we draw inspiration from recent advancements in speech self-supervised model, WavLM \cite{chen2022wavlm} and
introduce the usage of WavLM as a front-end feature extractor for the first time.
Attributed to the masked speech prediction and denoising training, WavLM has the potential to learn non-ASR features, including complex acoustic environments and speaker-related information
and had demonstrated improved performance on certain non-ASR tasks.
Additionally, we propose the Multi-Fusion Attentive (MFA) classifier based on the attentive statistics pooling layer \cite{asp}.
The MFA aggregates the output representations of WavLM to focus on features at different layers and time steps, thus facilitating the extraction of highly discriminative features.

% 再写文章架构
This paper is organized as follows.
Sec.\ref{sec:method} elaborates on the proposed detection system.
Sec.\ref{sec:experiments} provides a detailed description of the implementation and training details.
The results are discussed in Sec.\ref{sec:results}.
Finally, in Sec.\ref{sec:conclusion}, we summarize our methods and draw conclusions.

\section{Proposed Method}
\label{sec:method}

\begin{figure}[htbp]
    \centering
    \includegraphics[width=0.99\linewidth]{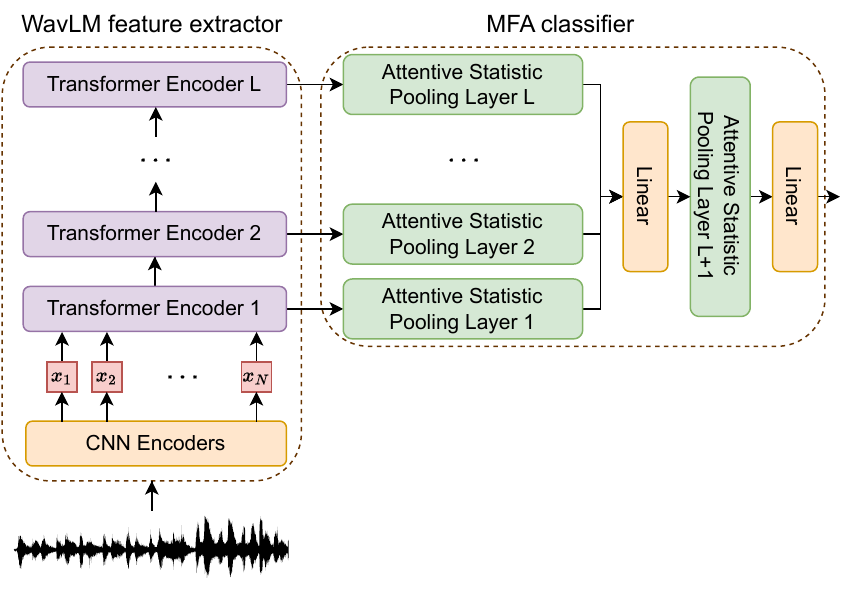}
    \caption{Pipeline of our proposed model. Left part: the WavLM model; Right part: the MFA Classifier.}
    \label{fig:framework}
\end{figure}

The overall framework of the model is illustrated in Fig.\ref{fig:framework}.
The input raw waveforms are fed into the WavLM model to obtain layers of frames of feature embedding.
All the embeddings are then passed through the Multi-Fusion Attentive (MFA) classifier to get the final prediction.
This section provides a detailed description of these modules.

\subsection{WavLM Model}
\label{subsec:wavlm}
WavLM is a speech self-supervised model that employs Wav2vec2 as its backbone.
It consists of a convolutional feature encoder and Transformer encoders.
The convolutional feature encoder converts the raw waveform into a feature sequence $\mathbf{X}=\{\mathbf{x}_1,\mathbf{x}_2,\cdots,\mathbf{x}_N\}$ of length $N$, where $N$ is the number of frames.
Transformer encoders are composed of several Transformer layers,
where the input for the first layer is the output features retrieved by the CNN encoder,
and the input for each subsequent layer is derived from the preceding layer.
The output representation of the $l$-th Transformer layer be denoted by $\mathbf{H}^l=\{\mathbf{h}_1^l, \mathbf{h}_2^l, \cdots, \mathbf{h}_N^l\}$.
Let $\mathcal{H}=\{\mathbf{H}^l\}^{L}_{l=1}$ be the set of output representations from all layers,
where $L$ is the number of Transformer Encoder layers.

The WavLM utilizes a masked speech denoising and prediction framework, where noise (or overlapping) is added to the input audio before masking.
For the masked frames, the model is trained to predict pseudo-labels.
% The model is trained to predict pseudo-labels for the masked frames.
The masked speech denoising allows WavLM to learn non-ASR features related to speaker characteristics and acoustic environments,
making it more suitable for our audio detection tasks, since fake speech often comprises many speaker-related artefacts.

\subsection{ASP-based MFA}
\label{subsec:aspc}

% Here we introduce our proposed ASP-based Multi-Fusion Attentive Classifier.
% We first describe the underlying mechanism of Attentive Statistics Pooling \cite{asp},
% followed by the presentation of our proposed classifier based on the ASP layer.

\subsubsection{Attentive Statistics Pooling}
\label{subsubsec:asp}

Attentive Statistics Pooling (ASP) \cite{asp} was originally used for extracting speaker embeddings and achieved excellent performance in speaker verification tasks \cite{desplanques20_interspeech}.
ASP is a pooling method that combines the advantages of both attention and statistic pooling.
% For our audio deepfake detection, 
Higher-order statistics can enhance speaker discriminability, and thus facilitate the detection of speaker-related artefacts in spoofed samples.
% Additionally, attention mechanisms highlight different positional features to contribute to the final output,
% such as assigning smaller weights to silent segments and larger weights to voiced segments.

Given a feature sequence $\mathbf{z}_t, \enspace t=1,\cdots,T$, ASP calculates the weighted mean $\mathbf{\mu}$ and standard deviation $\mathbf{\sigma}$ of the sequence as follows:
\begin{equation}
    \begin{aligned}
        \mathbf{\mu}    & = \sum_{t=1}^{T} \alpha_t \mathbf{z}_t,                                                         \\
        \mathbf{\sigma} & =\sqrt{\sum_t^T \alpha_t \mathbf{z}_t \odot \mathbf{z}_t-{\mathbf{\mu}} \odot {\mathbf{\mu}}},
    \end{aligned}
\end{equation}
where $\alpha_t$ is the attention weight of the $t$-th frame, and $\odot$ denotes the element-wise product.
Attention weights are computed for each frame through linear layers that output a scalar,
which is then normalized using a softmax function:
\begin{equation}
    \begin{aligned}
        e_t      & =\boldsymbol{v}^T f\left(\boldsymbol{W} \mathbf{z}_t+\boldsymbol{b}\right)+k, \\
        \alpha_t & =\frac{\exp \left(e_t\right)}{\sum_\tau^T \exp \left(e_\tau\right)},
    \end{aligned}
\end{equation}
where $\boldsymbol{v}$, $\boldsymbol{W}$, and $\boldsymbol{b}$ are learnable parameters of the modules.

\subsubsection{MFA Classifier}
\label{subsubsec:mfa}
% Chen_2022
Previous work \cite{donas_icassp, Chen_2022, pepino21_interspeech} has demonstrated that the intermediate feature of self-supervised model contains certain discriminative information.
According to \cite{yang21c_interspeech}, the authors have also indicated that by weighting the representations of different layers, self-supervised model has shown great potential in various speech downstream tasks.
To learn the optimal weight configuration,
% of different layers in self-supervised model, 
we propose the Multi-Fusion Attentive classifier based on the ASP layer.
The MFA classifier consists of the stacked time-wise ASP (T-ASP) layers and a single layer-wise ASP (L-ASP) layer.
The T-ASP layer is used to extract the time-level features from the hidden representations of different transformer layer.
% and the number of T-ASP layers is equal to the number of transformer layers of WavLM.
Each T-ASP computes a concatenated representation of the mean and standard deviation of the input sequence to form a single vector.
Finally, the L-ASP layer incorporates all the layer-level features to generate the output representation.

More specifically, given the output representations $\mathcal{H}$ from all $L$ layers of WavLM Transformer encoders, there are $L$ independent ASP layers that act on the time channel of the representations, computing the statistical pooling mean and variance via:
\begin{equation}
    \begin{aligned}
        \mathbf{r}^l & = \operatorname{concatenate}(\mathbf{\mu}^l, \mathbf{\sigma}^l) \\
                     & =  \operatorname{T-ASP}_{l}(\mathbf{H}^l), \quad l=1,\cdots,L,
    \end{aligned}
\end{equation}
Where the $\operatorname{T-ASP}_{l}$ denotes the T-ASP layer corresponding to the $l$-th layer of WavLM.
All the layer-level statistical pooling representations are then stacked to form a new channel, 
where an additional L-ASP layer is applied to perform attention calculation as
\begin{equation}
    \mathbf{o} = \operatorname{L-ASP}(\operatorname{concatenate}(\mathbf{r}^1,\mathbf{r}^2,\cdots,\mathbf{r}^L)).
\end{equation}

Output representation $\mathbf{o}$ is then fed into fully connected layers to obtain the final prediction.

% 实验

\section{Experiments}
\label{sec:experiments}
% Finally, reports the results and analysis.

\subsection{Datasets and metrics}
\label{subsec:datasets}

% We employ the training and developing sets from the ASVspoof 2019 LA  training set and development set \cite{todisco19_interspeech}, and evaluate our approach on the ASVspoof 2019 LA, ASVspoof 2021 LA, and ASVspoof 2021 DF evaluation set \cite{yamagishi21_asvspoof}.
% Since DF set contains a variety of bona fide and spoofed speech, and many trails come from mismatched domains, it makes the task challenging.
% We report min t-DCF and EER of our proposed systems.
% The evaluation metric mainly used in the experiment is the equal error rate (EER),
% which reflects the model's ability to distinguish genuine and fake trails, and a lower value indicates a better performance.
% Experiments are conducted on the ASVspoof 2019 Logical Access (LA) and DeepFake (DF) database \cite{todisco19_interspeech}.
% LA partition is divided into three subsets: training, development, and evaluation.
% We test our model on the ASVspoof 2019 LA, ASVspoof 2021 LA, and ASVspoof 2021 DeepFake (DF) evaluation set \cite{yamagishi21_asvspoof},
We focused on the logical access (LA) and speech deepfake (DF) partitions of the ASVspoof 2021 challenge \cite{yamagishi21_asvspoof}.
With no new training or development data being released, 
the ASVspoof 2021 challenge requires the use of the training and development partitions of the ASVspoof 2019 databases \cite{todisco19_interspeech}.
So we use ASVspoof 2019 LA train and dev sets for training, and evaluate our approach on the ASVspoof 2019 LA, ASVspoof 2021 LA, and ASVspoof 2021 DF evaluation set.

The 2021 LA and DF sets are similar to the 2019 LA data, but are intentionally more challenging: 
the LA evaluation data contains new trials for each speaker and both encoding and transmission artefacts,
and the DF evaluation data exhibits audio coding and compression artefacts.

We evaluate our model with two metrics: the minimum normalized tandem detection cost function (min t-DCF) \cite{kinnunen2018t} and equal error rate (EER).
The min t-DCF assesses the combined (tandem) performance of CMs and ASV whereas the EER reflects the standalone spoofing detection performance.

\subsection{Implementation Details}
\label{subsec:details}
% 包括
% 音频处理
% 1. 数据增强
% 2. 优化器参数
% 3. 采用 pytorch lightning 进行训练
% 4. 在 3090 上进行训练
In the pre-processing stage, the audio samples are pre-emphasized with a coefficient of 0.97
and then truncated or concatenated to a fixed length of approximately 4 seconds (64600 sample points).
We do not apply voice activity detection or any normalization to the audio samples.

Models are trained using Adam optimizer with $\beta=[0.9, 0.999]$.
We employ a step learning-rate decay scheduler to accelerate convergence.
% Hyperparameters may vary depending on whether or not to fine-tune the pre-trained frontend.
For the frontend WavLM fine-tuning, we set the initial learning rate at $3\times 10^{-6}$, which decays every $6000$ steps with a decay factor, or gamma, of $0.1$. The batch size is $4$.
% For the frontend fine-tuning, the initial learning rate is set to $3\times 10^{-6}$, with a step size of $6000$, gamma of $0.1$, and a batch size of $4$.
When the WavLM is fixed, the batch size is increased to $32$, the learning rate is set to $0.003$, and the step size and gamma are $3200$ and $0.5$, respectively.

% It should be noted that we do {\bf not} perform any hyperparameter searching.
The entire experiments are conducted on four NVIDIA GeForce RTX 3090 GPUs.
%  and the results are reproducible under the same random seed and GPU environments.
% The whole architecture is implemented using PyTorch Lightning \footnote{https://github.com/Lightning-AI/lightning}.
For each configuration, the model is trained for about 16,000 steps.
% We use the pre-trained WavLM Base and WavLM Large models provided by the authors official repository \footnote{https://github.com/microsoft/unilm/tree/master/wavlm}.
The WavLM model and MFA classifier are trained jointly.
For the WavLM model, we use the pre-trained weights provided by the authors official repository \footnote{https://github.com/microsoft/unilm/tree/master/wavlm} as the initialization parameters.
We use two different WavLM models: WavLM Base and WavLM Large.
% For the Large (Base) model, the dimension of the output representation is 768 (1024), and the number of Transformer layers is 24 (12).
% For the Large model, the dimension of the output representation is 1024, and the number of Transformer layers is 24.
The dimension of the hidden representation of the MFA classifier is set to match the dimension of the output representation,
and the number of T-ASP layers is equal to the number of transformer layers in WavLM.
% The two best-performing models on the validation set are chosen for testing, and the best EER performance on the test set is the final result.

\section{Results and Analysis}
\label{sec:results}

\subsection{Results on ASVspoof 2021 LA and DF evaluation set}
\label{subsec:test_results}

Table \ref{table:method_comparison} presents a comparison of our results and those of other models that use self-supervised front-ends on the ASVspoof 2021 LA and DF evaluation dataset.
As is shown, our approach with WavLM and MFA achieves the best performance on the DF set and a competitive result on the LA set.
To the best of our knowledge, it is the lowest reported EER on the DF evaluation set.

We explain the reasons behind our superior performance.
On the one hand, most existing approaches using self-supervised models adopt Wav2vec2 or its variations as the pre-trained front-end feature extractor.
However, since these models are trained by masked prediction, the contextualized representations from the model contain more information about the audio content than the speaker information.
Instead, the WavLM is pre-trained by masking speech denoising and prediction and is capable of learning more speaker-related information and complex acoustic environments,
such as speaker characteristics and diverse audio backgrounds.
Such information is particularly helpful for audio deepfake detection, as most spoofed speech typically contains speaker-related artefacts.
On the other hand, compared with \cite{wang22_odyssey} using a complex back-end model based on heterogeneous graph neural networks,
our proposed ASP-based MFA classifier is essentially the fully connected layers but can attend features of WavLM at different time and layer levels, leveraging the complementary information of audio features.
% whereas \cite{wang22_odyssey} uses a complex back-end model based on heterogeneous graph neural networks.

\begin{table}[htbp]
    \caption{Comparative Pooled EER (\%) results of our proposed method with other anti-spoofing systems based on the self-supervised model in the ASVspoof 2021 LA and DF evaluation set.}
    \begin{center}
        \begin{tabular}{cccccc}
            \hline
            \multirow{2}{*}{System}           & \multirow{2}{*}{Front-end} & \multicolumn{2}{c}{Pooled EER(\%)}              \\
            \cline{3-4}
                                              &                            & DF                                 & LA         \\
            \hline
            Wang et al. \cite{wang22_odyssey} & Wav2vec2-XLSR              & 5.44                               & 7.18       \\
            \hline
            Doñas et al. \cite{donas_icassp}  & Wav2vec2-XLS128            & 4.98                               & 3.54       \\
            \hline
            Wang et al. \cite{wang22_odyssey} & Wav2vec2-XLSR              & 4.75                               & 6.53       \\
            % \hline
            % Eom et al. \cite{eom22_interspeech} & Wav2vec2                                     & -                          & 4.92 \\
            \hline
            Tak et al. \cite{tak22_odyssey}   & Wav2vec2                   & 2.85                               & {\bf 0.82} \\
            \hline
            {{\bf Ours}}                      & WavLM-Large                & \bf{2.56}                          & 5.08       \\
            \hline
        \end{tabular}
        \label{table:method_comparison}
    \end{center}
\end{table}

\subsection{Results on ASVspoof 2019 LA evaluation set}
\label{subsec:la_results}
We further test our model on the ASVspoof 2019 LA evaluation set, results are shown in Table \ref{table:la_results}.
% According to the results, we achieve better performance compared with recent anti-spoofing systems. 
Our approach demonstrates a pooled min t-DCF of 0.0126 and an EER of 0.42\%.
% , which is a competitive result compared with recent anti-spoofing systems.
% Besides, our method features a relatively concise architecture without any additional modules like VIB.
% Results also show that our WavLM front-end outperforms most non-self-supervised front-end models, which is also the reason why we choose WavLM as the feature extractor.
Results show that our model achieve better performance compared with recent anti-spoofing systems,
% outperforms nearly all non-self-supervised architectures, 
demonstrating the effectiveness and superiority of our proposed method.

% \begin{table}[htbp]
%     \caption{Comparison with other anti-spoofing systems in the ASVspoof 2019 LA evaluation set, reported in terms of pooled min t-DCF and EER(\%).}
%     \begin{center}
%         \begin{tabular}{ccc}
%             \hline
%             System & min t-DCF & EER(\%) \\
%             \hline
%             Yang et al. \cite{yang2023comparative} & 0.0520 & 1.92 \\
%             \hline
%             Hua et al. \cite{hua2021towards} & 0.0481 & 1.64 \\
%             \hline
%             Tak et al. \cite{tak21_asvspoof} & 0.0335 & 1.06 \\
%             \hline
%             Jung et al. \cite{AASIST} & 0.0275 & 0.83 \\
%             \hline
%             Huang et al. \cite{huang2023discriminative} & 0.0176 & 0.52 \\
%             \hline
%             Ours & 0.0126 & 0.42 \\
%             \hline
%             Eom et al. \cite{eom22_interspeech} & 0.0107 & 0.40 \\
%             \hline
%         \end{tabular}
%         \label{table:la_results}
%     \end{center}
% \end{table}
\begin{table}[htbp]
    \caption{Comparison with other anti-spoofing systems in the ASVspoof 2019 LA evaluation set, reported in terms of pooled min t-DCF and EER(\%).}
    \begin{center}
        \begin{tabular}{ccc}
            \hline
            System                                      & min t-DCF & EER(\%) \\
            \hline
            Hua et al. \cite{hua2021towards}            & 0.0481    & 1.64    \\
            \hline
            Yang et al. \cite{yang2023comparative}      & 0.0360    & 1.21    \\
            \hline
            Zhang et al. \cite{zhang122021effect}       & 0.0368    & 1.14    \\
            \hline
            Tak et al. \cite{tak21_asvspoof}            & 0.0335    & 1.06    \\
            \hline
            Jung et al. \cite{AASIST}                   & 0.0275    & 0.83    \\
            \hline
            Huang et al. \cite{huang2023discriminative} & 0.0176    & 0.52    \\
            \hline
            Ours                                        & 0.0126    & 0.42    \\
            % \hline
            % Eom et al. \cite{eom22_interspeech} & 0.0107 & 0.40 \\
            \hline
        \end{tabular}
        \label{table:la_results}
    \end{center}
\end{table}

\subsection{Ablation Study}
\label{subsubsec:ablation_study}

Table \ref{table:ablation} describes the results of our ablation experiment on each component of the modified architecture.
% We compare different front-end models, whether to perform fine-tuning and different classifiers.
The performance deteriorates significantly with Wav2vec2.
It is also worth noting that when using WavLM-L as the front-end, our approach achieves an EER of 5.79\% even without fine-tuning.
This is a very competitive result compared with fine-tuned Wav2vec2.
The features of WavLM exhibit substantial resilience to variations in speech within the DF dataset,
resulting in a significant reduction in the EER.
% when compared to Wav2vec2 features.

As for the back-end classifier, results show that the ASP-based MFA classifier is beneficial.
The GAP classifier only utilizes final layer representations of WavLM, thereby exhibiting a constrained capacity to capture the self-supervised representations.
And the TN + FCs classifier applies simple time-wise normalization, 
%  by weighting the representations of each layer, 
which is less effective for higher speaker discriminability.
The two classifiers resulted in a relative degradation in performance of 136.6\% and 15.0\%, respectively.

\begin{table}[htbp]
    \caption{
        The ablation study intends to demonstrate the effectiveness of each component of the system.
            {\it ft}: fine tuning.
            {\it -L} and {\it -S}: large and small.
            {\it ASP}: attentive statistics pooling.
            {\it GAP}: global average pooling.
            {\it TN}: temporal normalization.
            {\it FCs}: fully connected layers.
    }
    \begin{center}
        \begin{tabular}{ccc}
            \hline
            Ablation                     & Configuration & Pooled EER \\
            \hline
            WavLM-L(ft) + MFA            & -             & {\bf 2.56} \\
            \hline
            \multirow{3}{*}{w/o WavLM-L} & Wav2vec2-S    & 11.68      \\
                                         & Wav2vec2-L    & 10.03      \\
                                         & WavLM-S       & 10.06      \\
            \hline
            \multirow{2}{*}{w/o ft}      & WavLM-S       & 10.81      \\
                                         & WavLM-L       & 5.79       \\
            \hline
            \multirow{2}{*}{w/o MFA}     & GAP           & 6.01       \\
                                         & TN + FCs      & 2.92       \\
            \hline
        \end{tabular}
        \label{table:ablation}
    \end{center}
\end{table}

\section{Conclusion}
\label{sec:conclusion}
In this paper, we focus on audio deepfake detection based on the speech self-supervised model.
We employ the self-supervised WavLM as the front-end feature extractor for the first time 
and propose a novel ASP-based Multi-Fusion Attentive classifier.
The multi-layer representations of WavLM are aggregated by the time-wise and layer-wise ASP, which can effectively capture the complementary information of audio features.
By jointly training the WavLM and MFA, we show that the proposed methods show competitive results on both the ASVspoof 2019 and ASVspoof 2021 datasets.
% achieve the best performance on the ASVspoof 2021 DF dataset
% and get competitive results on the LA set.

% 致谢
% \section{Acknowledgements}
% \label{sec:acknowledgements}
% We would like to express our gratitude to Mr. Zhang Zhan and Ms. Wang Jinyang for their insightful comments.

% \vfill\pagebreak
\bibliographystyle{IEEEbib}
\bibliography{refs}

\end{document}